\documentclass[fleqn,usenatbib]{mnras}

\usepackage{newtxtext,newtxmath}
\usepackage[T1]{fontenc}

\DeclareRobustCommand{\VAN}[3]{#2}
\let\VANthebibliography\thebibliography
\def\thebibliography{\DeclareRobustCommand{\VAN}[3]{##3}\VANthebibliography}

\usepackage{dcolumn, color, units, xspace, mathtools, physics, tensor, bm, lipsum, revsymb}
\usepackage[caption=false]{subfig}
\usepackage[normalem]{ulem}
\usepackage{import} % to import tex files as subsections within this documents 
\usepackage{multirow}
\usepackage{xspace}
\usepackage{acronym}

\usepackage{graphicx}	% Including figure files
\usepackage{amsmath}	% Advanced maths commands

\title[Systematic errors in searches for nanohertz gravitational waves]{Systematic errors in searches for nanohertz gravitational waves}

\author[Valentina Di Marco et al.]{
Valentina Di Marco,$^{1,2,3}$\thanks{E-mail: valentina.dimarco@monash.edu}
Andrew Zic,$^{3}$
Ryan M. Shannon$^{4,5}$
and Eric Thrane$^{1,2}$
\\
$^{1}$School of Physics and Astronomy, Monash University, Clayton VIC 3800, Australia\\
$^{2}$OzGrav: The ARC Center of Excellence for Gravitational Wave Discovery, Clayton VIC 3800, Australia\\
$^{3}$CSIRO, Space and Astronomy, PO Box 76, Epping, NSW 1710, Australia\\
$^{4}$Centre for Astrophysics and Supercomputing, Swinburne University of Technology, Hawthorn VIC 3122, Australia\\
$^{5}$OzGrav: The ARC Center of Excellence for Gravitational Wave Discovery, Hawthorn VIC 3122, Australia\\
}

\date{Accepted XXX. Received YYY; in original form ZZZ}

\pubyear{2024}

\begin{document}
\label{firstpage}
\pagerange{\pageref{firstpage}--\pageref{lastpage}}
\maketitle

\begin{abstract}
A number of pulsar timing arrays have recently reported preliminary evidence for the existence of a nanohertz frequency gravitational-wave background.
These analyses rely on detailed noise analyses, which are inherently complex due to the many astrophysical and instrumental factors that contribute to the pulsar noise budget. 
We investigate whether realistic systematic errors, stemming from misspecified noise models that fail to capture salient features of the pulsar timing noise, could bias the evidence for gravitational waves.
We consider two plausible forms of misspecification: small unmodeled jumps and unmodeled chromatic noise. 
Using simulated data, we calculate the distribution of the commonly used optimal statistic with no signal present and using plausibly misspecified noise models.
By comparing the optimal statistic distribution with the distribution created using ``quasi-resampling'' techniques (such as sky scrambles and phase shifts), we endeavor to determine the extent to which plausible misspecification might lead to a false positive.
The results are reassuring: we find that quasi-resampling techniques tend to \textit{underestimate} the significance of pure-noise datasets.
We conclude that recent reported evidence for a nanohertz gravitational-wave background is likely robust to the most obvious sources of systematic errors; if anything, the significance of the signal is potentially underestimated.
\end{abstract}


%%%%%%%%%%%%%%%%%%%%%%%%%%%%%%%%%%%%%%%%%%%%%%%%%%

%%%%%%%%%%%%%%%%% BODY OF PAPER %%%%%%%%%%%%%%%%%%

\section{Introduction}\label{introduction}
The pulsar timing community has recently reported evidence for the angular correlations expected from a gravitational-wave background \citep{Antoniadis_2022, Agazie_2023_GW,  Reardon_2023_gw, Xu_2023}. 
The dominant source of nanohertz-frequency gravitational waves is expected to be from the inspiralling of supermassive black hole binaries \citep{Rajagopal_1995, Phinney_2001, Wyithe_2003}. 
Other proposed sources include cosmic strings \citep{Damour_2000, Damour_2001, Sanidas_2013} and phase transitions in the early Universe \citep{Maggiore_2000, Maggiore_2001, Caprini_2010}. 
Recent evidence for a gravitational-wave background provides a unique opportunity to study these phenomena.

One of the challenges in detecting the gravitational-wave background is that multiple other processes imprint low-frequency gravitational wave signals in the timing residuals for each pulsar. 
What distinguishes the intrinsic noise of the pulsars from a gravitational-wave signal is the unique feature of the angular correlation in the arrival times between pulsar pairs. Specifically, the Hellings and Downs function \citep{Hellings_1983} describes the expected quadrupolar correlation between the pulsar pairs as a function of their angle of separation. 
If a gravitational-wave signal is present in the data, we would expect the correlation between pulsar pairs to follow the shape of the Hellings and Downs curve.

A gravitational-wave background is expected to induce fluctuations in the time of arrival of pulses from the pulsars in the arrays. 
These modulations are characterized by the timing residual cross-power spectral density $S_t(f)$, which is typically assumed to be a power law, i.e. $S_t(f)\propto f^{-\gamma}$, where $f$ is the frequency of the gravitational wave and $\gamma$ is the spectral index. 
The timing residual cross-power spectral density (units = $\unit[]{yr^3}$ \citep{Hobbs_2009} or units = $\unit[]{\mu s \;yr}$\footnote{These are equivalent to units of $\unit[]{ns^2 Hz^{-1}}$}) is simply related to other metrics including the redshift cross-power spectral density ($S_s$; units = $\unit[]{redshift^2\,Hz^{-1}}$), the strain power spectral density ($S_h$; units = $\unit[]{strain^2\,Hz^{-1}}$), the characteristic strain ($h_c(f)$; unitless), and the gravitational-wave strain amplitude ($A$; unitless).\footnote{For supermassive back holes inspiralling solely due to gravitational waves, these quantities are proportional to the frequency in the following way
\begin{align*}
    h_c &\propto f^{-2/3}\\
    S_h &\propto f^{-7/3}\\
    S_s &\propto f^{-7/3}\\
    S_t &\propto f^{-13/3} .
\end{align*}.}

To ensure reliable detection, pulsar timing data analyses perform thorough noise characterization and mitigation \citep{EPTA_pulsar_noise, Reardon_2023_noise, Agazie_2023_noise, Falxa_2023} to account for confounding signals. 
However, this analysis is complex, and there could be sources of noise that are not well understood or known. 
If the model for the time variations of the pulsar is misspecified, the estimate of the correlated gravitational-wave signal could be affected by unaccounted features in the noise \citep{Goncharov_2021a, Goncharov_2021B, Shannon_2016, Tiburzi_2016}.
For a review of model misspecification in the context of gravitational waves; see \citealp{wmf}.

Noise analyses in pulsar timing include contributions from many stochastic processes. 
These stochastic processes, typically described with a power spectrum, include not only the gravitational-wave signal, but also the stochastic spin-down variations in the pulsars, the variations in the electron column density due to the turbulent interstellar medium, and more. 
The temporally correlated red noise can be divided into achromatic and chromatic noise. Achromatic noise, also called timing noise or spin noise, does not depend on radio frequencies and is largely associated with irregularities in the rotation of the neutron star \citep{Shannon_2010}.   
Strongly chromatic timing perturbations include contributions from plasma dispersion, Faraday rotation, and interstellar scattering \citep{Cordes_2010}. 
%The strongest source of red noise is believed to be the variation in the dispersion measure \citep{Keith_2013}. 
For a comprehensive study of noise in pulsar timing, see \citealp{Goncharov_2021a} and \citealp{ Lentati_2016}. 

One of the statistics used in gravitational wave detection analyses is the optimal statistic \citep{Anholm_2009, Demorest_2013, Siemens_2013, Vigeland_2018, Sardesai_2023, Hazboun_2023}. 
The optimal statistic is a frequentist statistic that estimates the power of the quadrupolar gravitational wave-correlated signal. 
It is derived by taking the ratio of the likelihood of a model with an angularly correlated gravitational-wave background and a model with only noise. 
It can also be derived as a weighted average of the correlation coefficients between pulsar pairs \citep{Agazie_2023_GW}. 
The optimal statistic can be determined for the quadrupolar signal of gravitational waves, but also for dipolar and monopolar signals. 
While detection of a quadrupolar signal within the dataset serves as compelling evidence for the presence of a gravitational wave, a dipolar signal in the data is typically associated with solar system ephemeris errors, whereas a monopolar signal can arise due to time-standard errors affecting all pulsars equally \citep{hobbs_2012, Tiburzi_2016, Hobbs_2019}.

The statistic used by various collaborations to assess the significance of detection is the optimal statistic signal-to-noise ratio (SNR) (see, for example, \citealp{Agazie_2023_GW}). 
This statistic is used to calculate the $p$-values by defining the distribution of the SNR in the absence of a gravitational-wave signal. 
However, it is not common practice in pulsar timing data analysis to take the SNR at face value so that, for example, SNR=5 implies a five-sigma $p$-value ($p \approx 3\times 10^{-7}$).
This is because a naive interpretation of the optimal SNR is only as reliable as the noise models used to calculate the optimal statistic. 
Therefore, collaborations use quasi-resampling methods such as sky scrambles and phase shifts to spoil the gravitational-wave correlations in the data while leaving the noise unchanged \citep{Cornish_2016, Taylor_2017}. 

Sky scrambling is designed to generate noise realizations by randomly redistributing each pulsar in the sky. 
This process is meant to eliminate any gravitational-wave correlations from the resulting scrambled data while retaining various noise artifacts. 
In the phase-shift approach, a new complex phase is randomly assigned to each pulsar strain measurement spoiling gravitational-wave-induced correlations while preserving the character of the noise.

In previous work \citep{Di_Marco_2023}, we evaluated how these tools are currently used by PTA collaborations and showed that the quasi-resampling methods do not produce a large number of statistically independent noise realizations---unlike the time slides used in ground-based gravitational-wave observatories \citep{Was_2009, Usman_2016, Capano_2017}.
We showed that it is still possible to derive reliable false-alarm probabilities using highly correlated noise realizations from sky scrambles and phase shifts---subject to the constraint that the noise models are adequately specified.
However, we presented a contrived example to show how misspecified noise could in theory yield a false-positive detection.

In this paper, we determine the extent to which \textit{realistic} noise misspecification affects the gravitational wave background estimation. 
We focus on two specific ways in which the noise models might be misspecified, which we regard as especially likely.
First, we consider the possibility that the data contain small unmodeled instrumental timing offsets (referred to as ``jumps’’) \citep{Kerr_2020}, which are known to occur, for example, when switching the back-end instruments of the telescope \citep{sarkissian_2011}. 
Second, we consider unmodeled chromatic noise present in some pulsars \citep{Cordes_2010}.
The remainder of this paper is structured as follows. 
In Section \ref{simulations}, we describe the procedure for simulating misspecified noise due to unmodeled jumps and/or chromatic noise. 
In Section \ref{results} we present results showing how background estimation is affected by this unmodeled noise.
We provide concluding remarks in Section \ref{conclusions}.

\section{Simulations and methods}\label{simulations}
In this section, we describe the simulations of the residuals of the pulsar timing array and detail the misspecifications and the methodology used for evaluating the detection statistic and the corresponding p-values.

Simulations are carried out with \texttt{ PTAsimulate} \footnote{(https://bitbucket.org/psrsoft/ptasimulate)}, a package that allows flexible and programmatic injection of signals into pulsar timing residuals.
The detection statistic (SNR), and the phase shift evaluation are performed using the package \texttt{Enterprise Extensions} \footnote{(https://github.com/nanograv/enterprise\_extensions)}, while sky scrambles are executed using our own code.
Parameter estimation has been performed with \texttt{Enterprise}. Both \texttt{Enterprise} and \texttt{Enterprise Extensions} are Python packages developed by the Nanograv Collaboration \citep{McLaughlin_2013}.

For both unmodeled jumps and chromatic noise analyses, we perform four studies, each with a different model.
\begin{itemize}
    \item The first model includes only white noise in the calculation of the SNR and quasi-resampling methods and using fixed parameters.
    \item The second model includes only white noise, but the parameters are free.
    \item The third model includes the intrinsic red noise (timing noise) of each pulsar with estimated parameters.
    \item The fourth model includes a common red process with estimated parameters.
\end{itemize}
These four models are summarized in Table \ref{table:Models}.
The first model is the least realistic because no real analyses use fixed parameter values.
However, we include it in order to see how a misspecified source of noise behaves when it is not absorbed into other quantities during parameter estimation.
The other three models are more realistic.
We use these models to evaluate whether the unmodeled noise from jumps and chromatic noise gets absorbed into the estimation of the white noise (model 1), intrinsic red noise (model 2), or common red noise (model 3).

The three estimated white noise parameters are \texttt{ECORR}, \texttt{EFAC} ND \texttt{EQUAD}  as defined in \citealp{Arzoumanian_2020}. 
\texttt{ECORR} or \textit{Extra CORRelated} white noise, accounts for the pulse phase jitter, which is an effect due to the folding of a finite number of pulses in each observing epoch leaving some residual shape fluctuation with respect to the profile template. 
\texttt{EFAC} or \textit{Extra FACtor}, is a multiplicative correction factor that accounts for systematic uncertainties causing the error bars of the time of arrivals to be overestimated or underestimated. 
\texttt{EQUAD} or \textit{Extra QUADrature}, adds white noise in quadrature and accounts for additional white noise Gaussian processes affecting the time of arrival the same way.
For example, if the timing residuals show a spread that is larger than the error bars of the time of arrivals, it is likely that there exists an additional uncertainty that we can model with this parameter.

In the fixed noise analysis, these parameters take the following values: \texttt{EFAC}=1, \texttt{EQUAD}=-8, \texttt{ECORR}=-8.
The intrinsic red noise is modeled as a power law, and the number of Fourier frequencies used is determined by the time span of each pulsar over the highest fluctuation frequency modeled, which in this case is 240 days.
For the common red noise, we use a model with a power-law spectrum, no inter-pulsar correlation, and a spectral index free to vary.
Since \texttt{Enterprise} and \texttt{Enterprise Extensions} require a common red signal for gravitational-wave searches using the optimal statistic, we add a common signal with a small amplitude: $10^{-20}$.

\begin{table}
\centering
\begin{tabular}{ |p{1.2cm}|p{1.2cm}|p{1.6cm}|p{2.6cm}|}
    \hline
    \multicolumn{4}{|c|}{Models used for SNR and quasi-resampling} \\
    \hline
    1 & 2 & 3 & 4\\
    \hline
    WN & WN & WN + RN & WN + RN + CRN\\
    fixed & free & free & free \\
    \hline
    \end{tabular}
    \caption{Model composition for the three analysis in the paper. The abbreviation WN stands for white noise, RN stands for intrinsic red noise, and CRN stands for common red noise. In Model 1 the parameters are fixed, while in all the other models the parameters are free.}
\label{table:Models}
\end{table}

\subsection{Unmodeled jumps}
In this paper, we focus on small jumps of instrumental origin and not of astrophysical nature. We seek to understand if these small jumps, if unmodeled, have an effect on the $p$-values of the SNR.

We perform 1000 simulations for the 30 pulsars used in PPTA DR3 \citep{zic_2023} of residual errors in three radio frequency bands ($\unit[600]{MHz}$, $\unit[1400]{MHz}$, and $\unit[3100]{MHz}$) in such a way that the root mean square (RMS) error is comparable to the whitened residuals as shown in Figure~8 of the PPTA DR3 noise analysis paper \citep{Reardon_2023_noise}.
We use a cadence of 25 days for 21.9 years of data.
Since our aim is to test the effect of misspecifications on a pure noise dataset, the simulations are performed in the absence of a gravitational wave signal.

In these simulations, we inject jumps with maximum amplitudes of 0.4, 4, 40 and $\unit[400]{ns}$.
The amplitudes of the 400 ns jumps are sampled from a uniform distribution with support [0, 400]. The max amplitudes are then subsequently proportionally reduced for the 40, 4 and $\unit[0.4]{ns}$ amplitudes.
The arrival times of the jumps are Poisson-distributed with an expected occurrence rate of $\unit[1]{yr^{-1}}$. 

Since the jumps are of instrumental origin, we assume that their arrival times and amplitudes are the same for every pulsar in the array and are only present at the $\unit[3100]{MHz}$ radio frequency, mimicking the effects observed in the Parkes single receivers in the past \citep{Manchester_2013, Arzoumanian_2015, Kerr_2020}. 
Jumps with amplitudes on the order of $\unit[40]{ns}$ or smaller are potentially too small to be detected in the residuals but could be a source of measurable inaccuracies.

For each simulation, we calculate the detection statistic without modeling these jumps and run 5000 sky scrambles and 5000 phase shifts with the same model used for the detection statistic. 
We then calculate the estimated $p$-values for this statistic for sky scrambles and phase shifts and plot them against the true $p$-values.

\subsection{Chromatic noise}
We repeat a similar analysis for the unmodeled chromatic noise, which could be present in some pulsars \citep{Goncharov_2021B, Reardon_2023_noise} but might not be accounted for in the noise analysis conducted by some of the collaborations that prefer a less complex model with fewer parameters.
Here, we focus on a general form of chromatic noise such that
\begin{equation}
    \Delta_{\text{CH}} \propto \nu^{-\chi}
\end{equation}
where $\Delta_{\text{CH}}$ is the induced delay in the observed signal and the chromaticity index $\chi$ can take any value.
In the PPTA DR3 noise analysis paper \citep{Reardon_2023_noise}, these forms of chromatic noise have been identified in seven pulsars, with indices ranging between $\approx 0.8-5.2$. 
As we did for the jump analysis, we investigate whether this unmodeled noise can have an effect on the SNR $p$-values.

We perform 1000 simulations of timing residuals for the 30 PPTA DR3 pulsars. As before, we simulate three radio frequency bands (600 \unit[]{MHz}, 1400 \unit[]{MHz}, and 3100 \unit[]{MHz}), we use a cadence of 25 days for 21.9 years of data.
We incorporate chromatic noise specifically for the seven pulsars where it is detected, following the analysis outlined in the PPTA DR3 noise analysis paper. In our simulation of chromatic noise, we utilize the parameters specified in Table~1 of \citealp{Reardon_2023_noise} for $\gamma^{\text{Chr}}$ and $\log_{10} A^{\text{Chr}}$ (see Table \ref{table:Chr_noise_param}), and the parameters specified in Table~2 of \citealp{Goncharov_2021B} for five of the seven pulsars.
As for the jumps analysis, for each simulation, we calculate the detection statistic without modeling the chromatic noise and run 5000 sky scrambles and 5000 phase shifts with the same model used for the detection statistic. 

\begin{table}
\centering
\begin{tabular}{ |p{2cm}|p{1.5cm}|p{2cm}| }
    \hline
    \multicolumn{3}{|c|}{Chrmoatic noise parameters} \\
    \hline
    PSR & $\gamma^{\text{Chr}}$ & $\log_{10} A^{\text{Chr}}$\\
    \hline
    J0437$-$4715 & 3.0 & -14.4\\
    J0613$-$0200 & 5.2 & -15.8\\
    J1017$-$7156 & 0.8 & -13.7\\
    J1045$-$4509 & 3.4 & -14.2\\
    J1600$-$3053 & 1.5 & -13.8\\
    J1643$-$1224 & 0.8 & -13.2\\
    J1939$+$2134 & 1.2 & -13.9\\
    \hline
    \end{tabular}
    \caption{
    Chromatic noise parameters for the seven pulsars identified with chromatic noise in the PPTA DR3 noise analysis.
    }
\label{table:Chr_noise_param}
\end{table}

\section{Results}\label{results}

\subsection{Unmodeled jumps analysis}
\begin{figure}
\includegraphics[clip,width=\columnwidth]{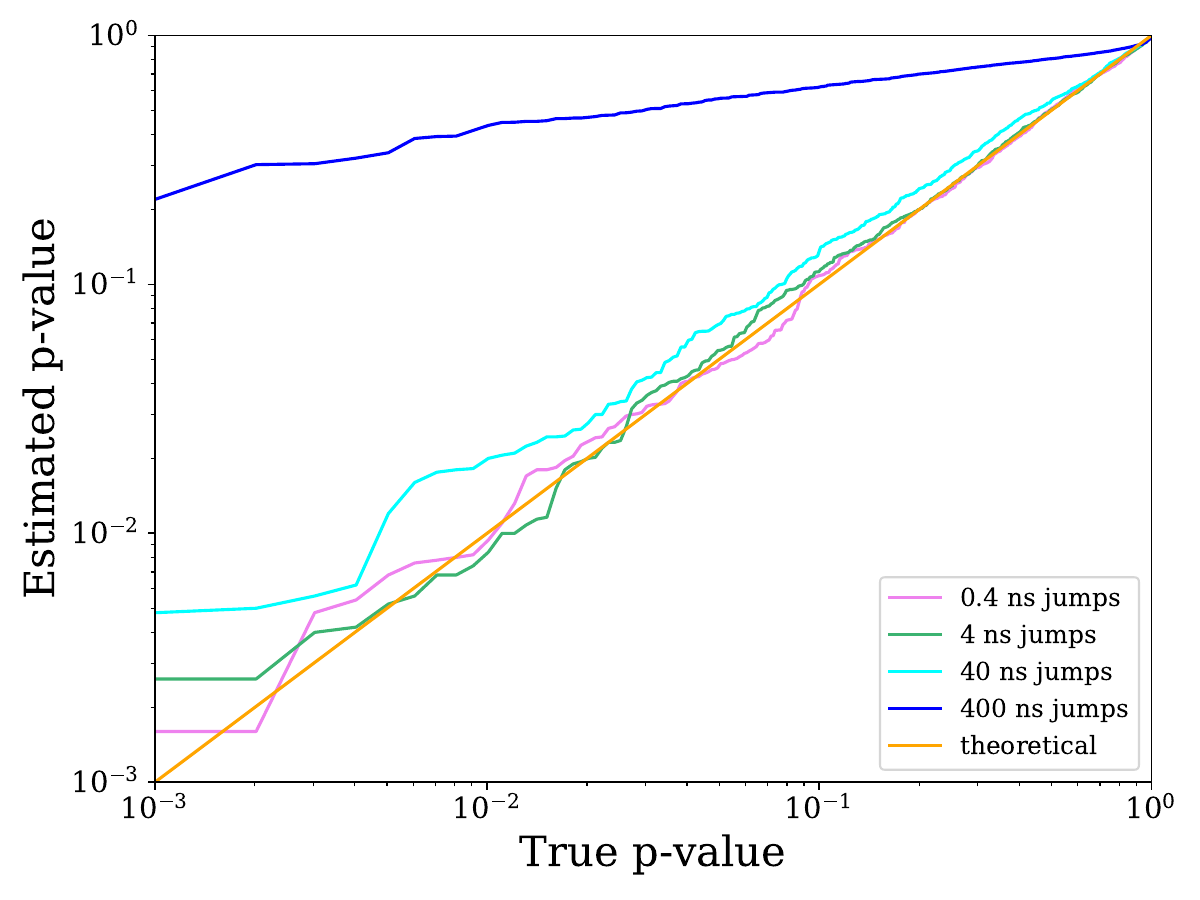}
\centering
\caption{
Phase shifts estimated $p$-value versus the true $p$-value for simulated data with unmodeled jumps. 
The orange line shows the expected trend, while the other lines show $p$-values for simulations with unmodeled jumps of different amplitudes from 0.4 \unit[]{ns} to 400 \unit[]{ns}. 
The different colours show results for the different jump sizes.
}
\label{fig:jumps_phase_shifts_god}
\end{figure}

\begin{figure}
\includegraphics[clip,width=\columnwidth]{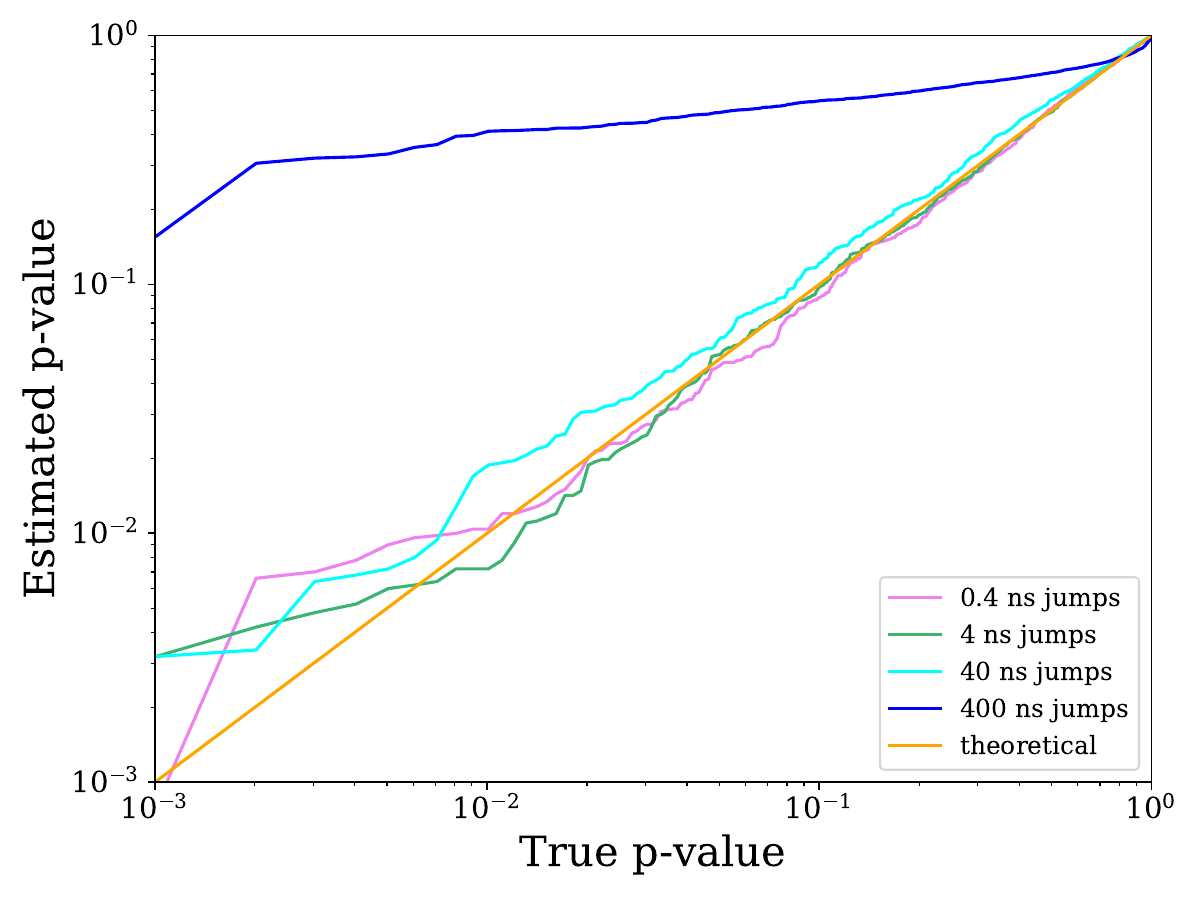}
\centering
\caption{Sky scrambles estimated $p$-value versus the true $p$-value for simulated data with unmodeled jumps. 
The orange straight line shows the expected trend, while the other lines show $p$-values for simulations with unmodeled jumps of different amplitudes from 0.4 \unit[]{ns} to 400 \unit[]{ns}. 
The different colours show the results for different jump sizes.
}
\label{fig:jumps_sky_scrambles_god}
\end{figure}

The effect of unmodeled jumps on phase-shift $p$-values is shown in Fig.~\ref{fig:jumps_phase_shifts_god} while Fig.~\ref{fig:jumps_sky_scrambles_god} shows the results for sky scrambles. 
This result is achieved using Model 1 in Table \ref{table:Models}. 
We find that the estimated $p$-values are greater than the true $p$-values for unmodeled jumps having amplitudes of $\sim 400$ ns and $\sim 40$ ns.
There is no discernible change when we use smaller maximum amplitudes. A similar result is achieved with Model 2.

We expect the presence of unmodeled jumps would induce a monopole signal, which could be a giveaway of this kind of misspecification. 
Therefore, for all simulations, we calculate the SNR monopolar signal and find that the distribution of this statistic has a nonzero mean (15.5) only for jumps with an amplitude of $\sim 400$ \unit[]{ns}. 
It follows that jumps with an amplitude of $\sim 40$ ns would have an effect on the detection statistic, but would not induce a detectable monopole.
When including intrinsic red noise in the models and estimating its parameters (Model 3) the noise due to the unmodeled jumps gets absorbed into the red noise parameters, so that jumps do not significantly influence $p$-values, at least for $p$-values as small as $10^{-3}$.

If the models also contain a common red noise (Model 4) the red noise effect due to the presence of unmodeled jumps is partially absorbed in the common red noise part of the model instead of the timing noise part. 
This is the most realistic scenario, as in their analysis the collaborations always include a common process.
In this case, the $p$-values are overestimated for all amplitudes, but this effect is more visible for the curve related to jumps with the highest amplitude ($\sim400$ ns).
This issue of characterizing a common red noise has been recently addressed \citep{Zic_2022, Goncharov_2021a}. Both papers show with simulated data how a spurious common red spectrum can arise from misspecified noise.
The interested reader can find the plots of the results for the analyses carried out with Models 2, 3 and 4 in Appendix \ref{Plots}.

\subsection{Unmodeled chromatic noise analysis}
\begin{figure}
    \centering
\includegraphics[clip,width=\columnwidth]{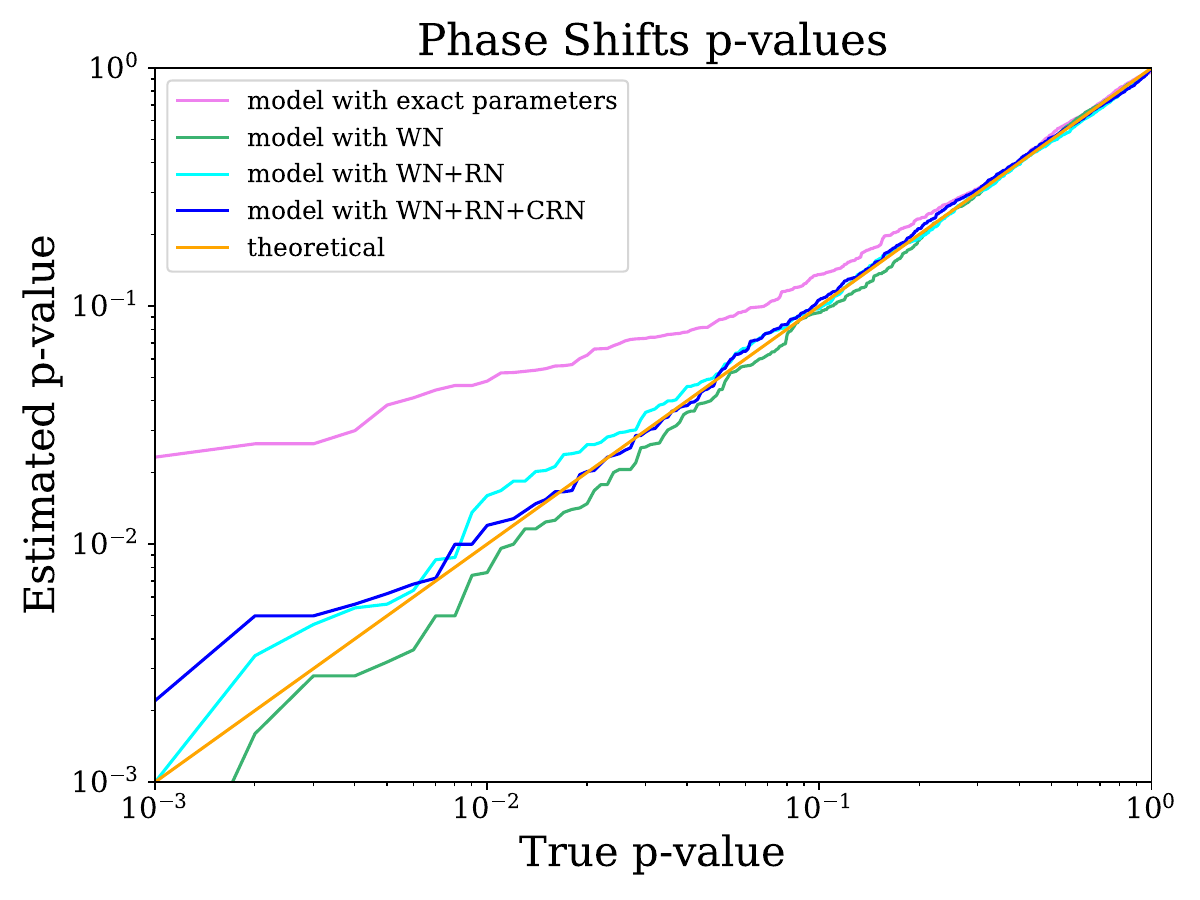}
    \caption{
    The estimated p-value versus the true p-value for simulated data with unmodeled chromatic noise for phase shifts. The orange straight line shows the expected trend while the other lines shows $p$-values for 3 models.}
    \label{fig:phase_shifts_chromatic}
\end{figure}

\begin{figure}
    \centering
\includegraphics[clip,width=\columnwidth]{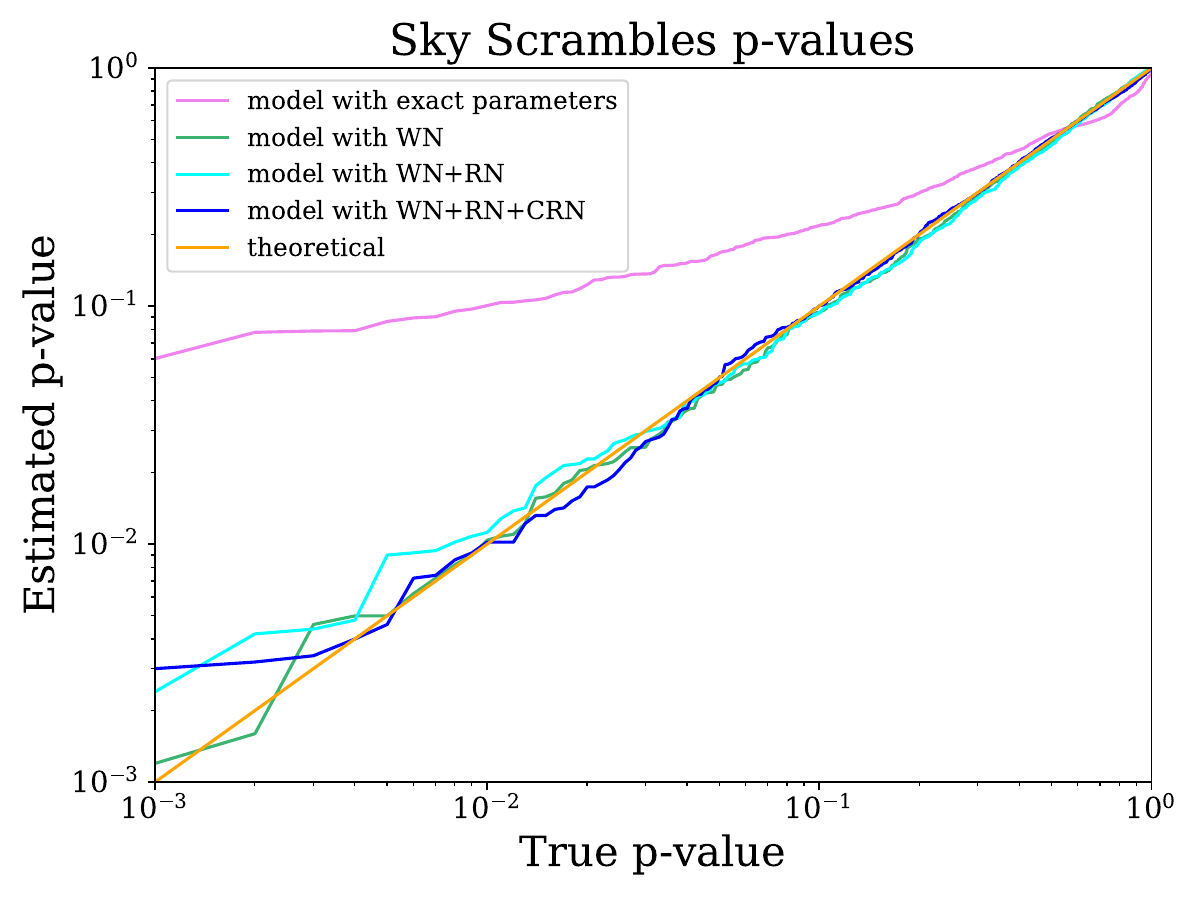}
    \caption{
    The estimated p-value versus the true p-value for simulated data with unmodeled chromatic noise for phase shifts. The orange straight line shows the expected trend while the other lines shows $p$-values for 3 models.}
    \label{fig:sky_scrambles_chromatic}
\end{figure}

The effect of unmodeled chromatic noise is shown in Fig.~\ref{fig:phase_shifts_chromatic} for phase shifts and in Fig.~\ref{fig:sky_scrambles_chromatic} for sky scrambles. 
We compare the four different models in Table \ref{table:Models}.
Similarly to the jump study, when we use a model that matches the simulations while not modeling the chromatic noise (Model 1), the $p$-values are overestimated. 
The analysis carried out with Model 2 shows that this misspecification gets absorbed in the white parameters, as only the model with fixed white noise parameters is clearly affected by the presence of unmodeled chromatic noise. Since no monopole is expected to be present, the more realistic models with estimated parameters are not affected by the misspecification.

\section{Discussion and Conclusions}\label{conclusions}
In this work, we utilize simulations to investigate how oversimplifications in noise modeling could potentially impact the detection of a nanohertz gravitational wave background. 
We examine two types of noise misspecification that we believe can be present in PTAs studies: unmodeled small jumps of instrumental origin and unmodeled chromatic noise present in only some pulsar. 
Our initial concern and motivation for this work was centered on the possibility that these inaccuracies in the noise assessments could lead to false positive detections. 
Instead, we find that these two forms of misspecified noise lead to a conservative overestimation of the $p$-value.

We observe the presence of a monopole in the data that contains jumps with amplitudes significant enough to be detectable in our residuals. 
However, smaller jumps, of the order of 40 ns, might not be easily detectable in the residuals and not produce a detectable monopole but would have an effect on the estimations of $p$-values.
We also find that, during parameter estimation, the noise from the unmodeled jumps is absorbed in the intrinsic red-noise parameters, while the unmodeled chromatic noise is absorbed in the white-noise parameters.
In the case of jumps, this means that in a realistic analysis in which one estimates the parameters for the full model, the $p$-values are overestimated, again resulting in a conservative detection statement. 
This also results in phase shifts that exhibit greater robustness to noise analysis inaccuracies in the context of jumps and clock errors.

In this paper, we only analyze two forms of realistic misspecifications that we believe could be present in PTA analyses. 
It is encouraging to find that quasi-resampling methods provide conservative estimations of $p$-values for the two forms of misspecification that we tested; however, it is still possible that we have failed to consider some form of misspecification that produces qualitatively different behaviors.

Beyond biasing detection statistics, misspecification may impact the recovered properties of the GW signal. This possibility has been raised, for example, in \citealp{Reardon_2023_gw, Reardon_2023_noise}, and may also contribute to slight discrepancies in the properties of the common signal reported across \citep{Agazie_2023_GW, EPTA_GW, Reardon_2023_gw}. 
These apparent discrepancies motivate further study into the effect of misspecification on gravitational-wave detection.

\section*{Acknowledgements}
We acknowledge and pay respects to the Elders and Traditional Owners of the land on which this work has been performed, the Bunurong, Wadawurrong and
Wurundjeri People of the Kulin Nation and the Wallumedegal People of the Darug Nation.
The authors are supported via the Australian Research Council (ARC) Centre of Excellence CE170100004.
E.T. is supported through ARC DP230103088.
V.D.M. receives support from the Australian Government Research Training Program. 
R.M.S. acknowledges support through ARC Future Fellowship FT190100155. 
This work was performed on the OzSTAR national facility at Swinburne University of Technology. The OzSTAR program receives funding in part from the Astronomy National Collaborative Research Infrastructure Strategy (NCRIS) allocation provided by the Australian Government.

%%%%%%%%%%%%%%%%%%%%%%%%%%%%%%%%%%%%%%%%%%%%%%%%%%
\section*{Data Availability}

No data was used in the analysis.



\bibliographystyle{mnras}
\bibliography{refs} % if your bibtex file is called example.bib

%\begin{thebibliography}{99}
%\end{thebibliography}

%%%%%%%%%%%%%%%%%%%%%%%%%%%%%%%%%%%%%%%%%%%%%%%%%%

%%%%%%%%%%%%%%%%% APPENDICES %%%%%%%%%%%%%%%%%%%%%

\appendix

\section{Plots for Models 3 and 4 of the jumps analysis} \label{Plots}

\begin{figure*}%
    %\subfloat[Model with fixed parameters - WN only\label{fig:PS-p-val-true}]{%
        %\includegraphics[width=0.5\columnwidth]{phase_shifts_god.pdf}
    %}\hfill
    \subfloat[Model with estimated parameters - WN only\label{fig:PS-p-val-WN}]{%
        \includegraphics[width=1\columnwidth]{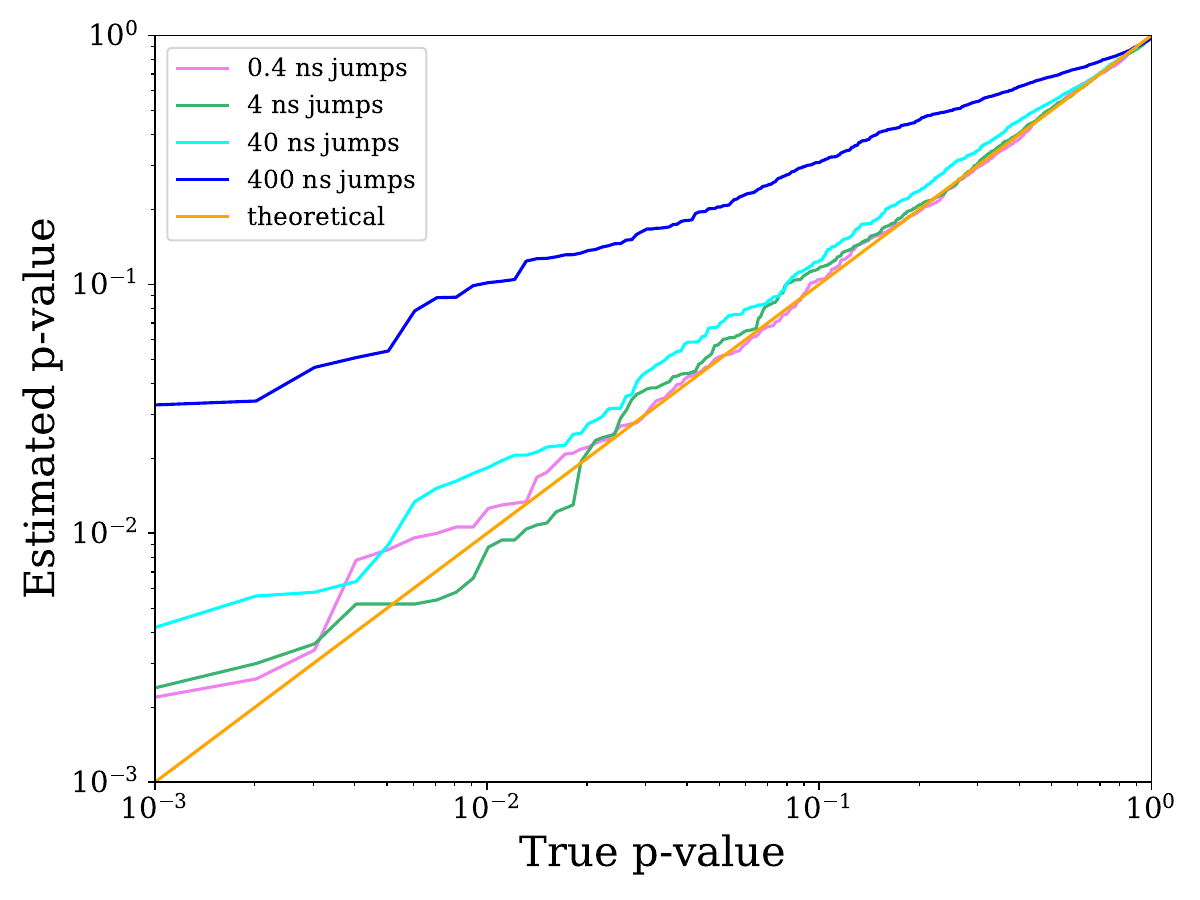}
    }\\
    \subfloat[Model with estimated parameters - WN + RN\label{fig:PS-p-val-WN-RN}]{%
        \includegraphics[width=1\columnwidth]{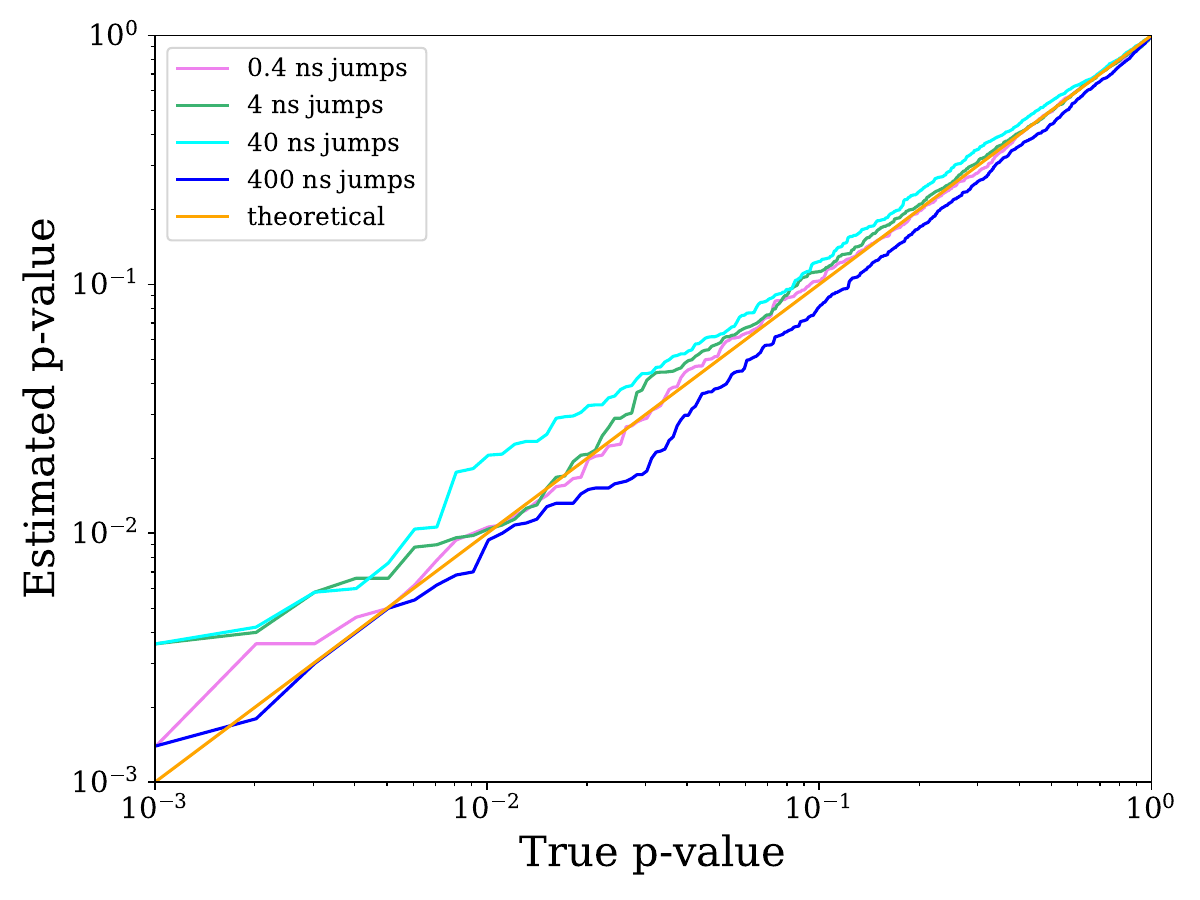}
    }\hfill
    \subfloat[Model with estimated parameters - WN + RN + CRN\label{fig:PS-p-val-WN-RN-CRN}]{%
    \includegraphics[width=1\columnwidth]{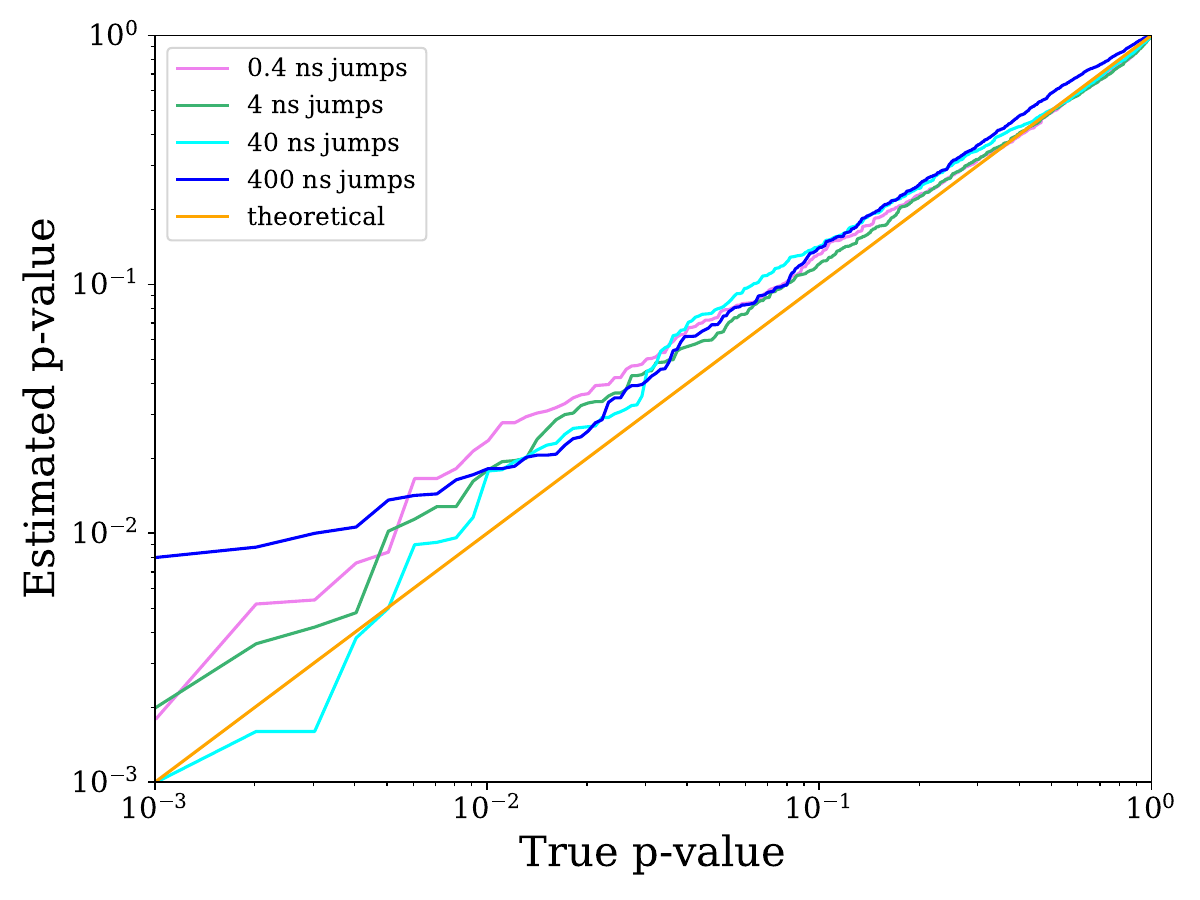}
    }
    \caption{Phase shifts estimated p-value versus the true p-value for simulated data with unmodeled jumps. The orange straight line shows the expected trend, while the other lines show $p$-values for simulations with unmodeled jumps of different amplitudes from 0.4 ns to 400 ns.}
\label{fig:jumps_phsae_shifts}
\end{figure*}

\begin{figure*}%
    %\subfloat[Model with fixed parameters - WN only\label{fig:SS-p-val-true}]{%
        %\includegraphics[width=0.5\columnwidth]{sky_scrambles_god.pdf}
    %}\hfill
    \subfloat[Model with estimated parameters - WN only\label{fig:SS-p-val-WN}]{%
        \includegraphics[width=1\columnwidth]{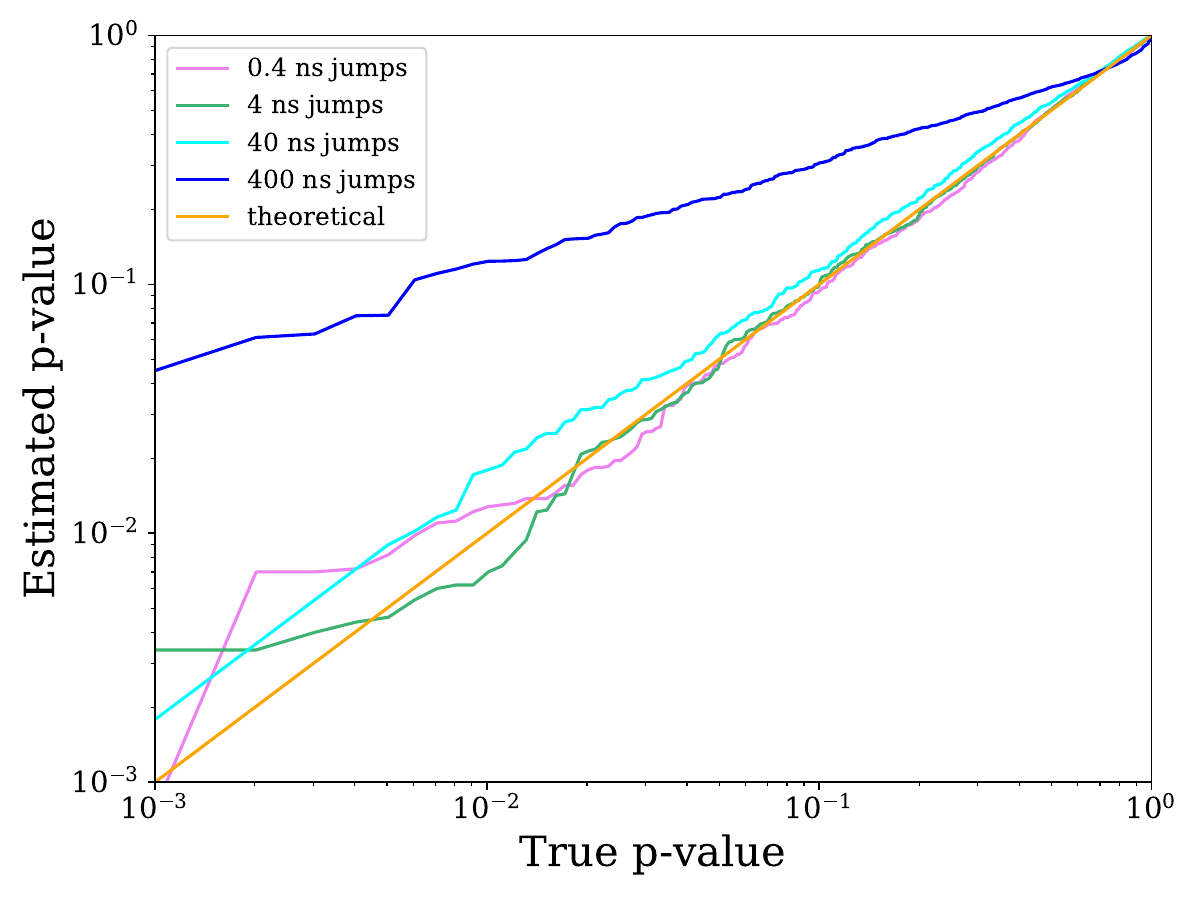}
    }\\
    \subfloat[Model with estimated parameters - WN + RN\label{fig:SS-p-val-WN-RN}]{%
        \includegraphics[width=1\columnwidth]{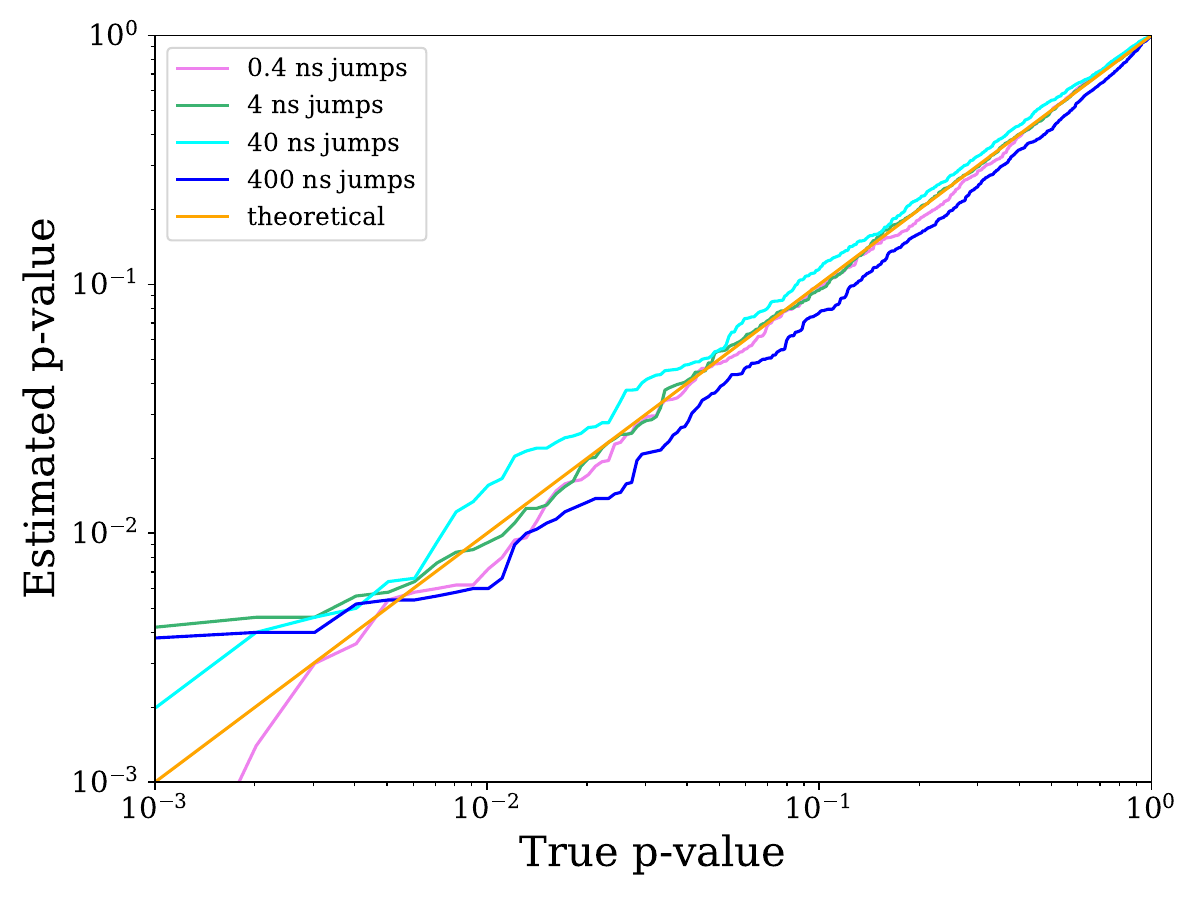}
    }\hfill
    \subfloat[Model with estimated parameters - WN + RN + CRN\label{fig:SS-p-val-WN-RN-CRN}]{%
        \includegraphics[width=1\columnwidth]{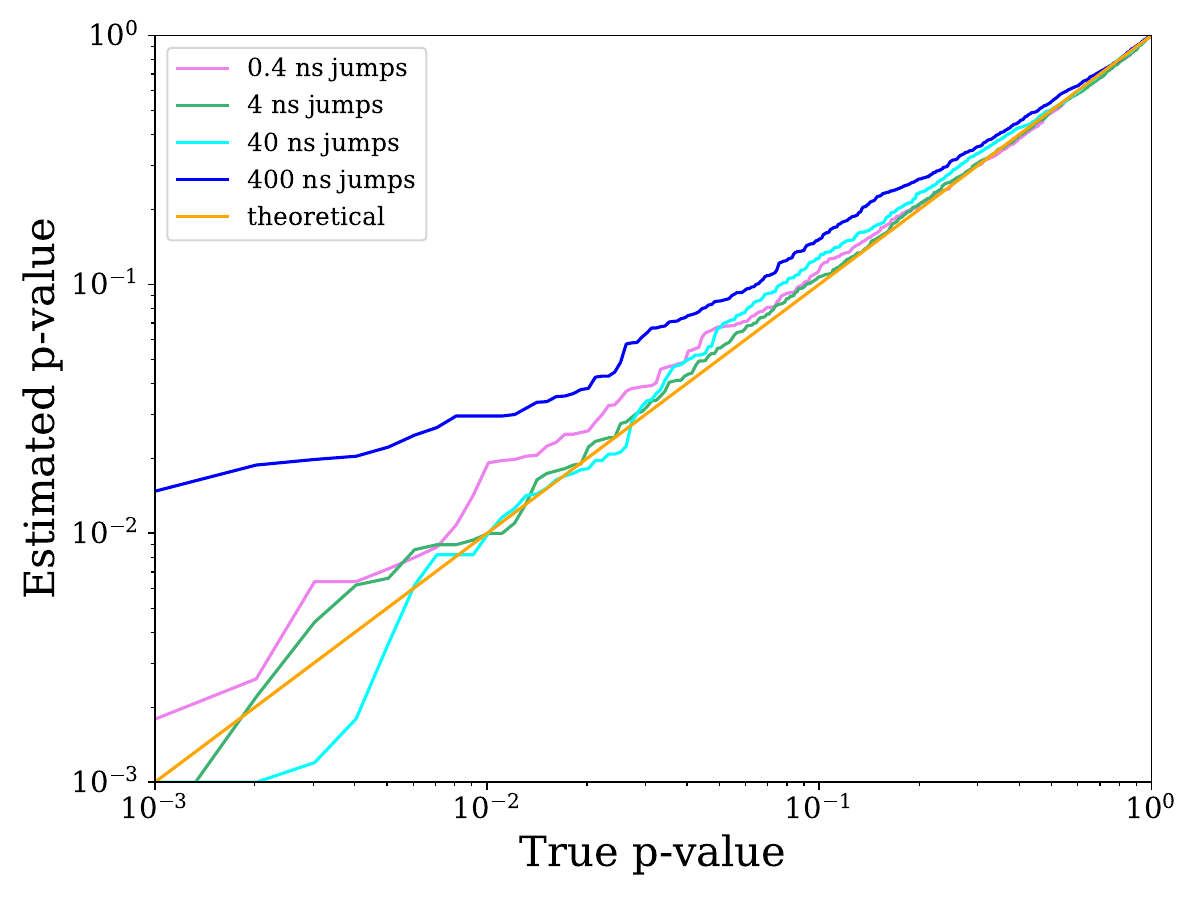}
    }
    \caption{Sky scrambles estimated p-value versus the true p-value for simulated data with unmodeled jumps. The orange straight line shows the expected trend, while the other lines show $p$-values for simulations with unmodeled jumps of different amplitudes from 0.4 ns to 400 ns.}
\label{fig:jumps_sky_scrambles}
\end{figure*}

%%%%%%%%%%%%%%%%%%%%%%%%%%%%%%%%%%%%%%%%%%%%%%%%%%

% Don't change these lines
\bsp	% typesetting comment
\label{lastpage}
\end{document}